\documentstyle[12pt]{article}

\title{ Comment on the Adiabatic Condition }

\author{  A.C. Aguiar Pinto$^{(1)}$,  M.C. Nemes$^{(2)}$,  \\
    J.G. Peixoto de Faria $^{(2)}$   and    M.T. Thomaz$^{(1)}$ 
\thanks{Corresponding author: 
        Dr. Maria Teresa Thomaz;
 R. Domingos S\'avio Nogueira Saad n.$\!\!^{\rm o}$ 120  apto 404, 
 Niter\'oi, R.J., 24210-310, BRAZIL
 Phone: (21) 620-6735, Fax: (21)620-3881;  
{\it E-mail}: mtt@if.uff.br. } 
 \\
\\
\baselineskip =10pt
{ \small \it $^{(1)}$ Instituto de F\'\i sica - Univ. Federal
                      Fluminense
\vspace{-0.2cm}}\\
{\small \it Av. Gal. Milton Tavares de Souza s/n.$\!\!^\circ$, 
                \vspace{-0.2cm} }\\
{ \small \it CEP: 24210-340, Niter\'oi, R.J.,  Brazil }  \\
{\small \it $^{(2)}$Departamento de F\'{\i}sica, ICEX,} 
\vspace{-0.2cm}\\
\vspace{-0.2cm}{\small \it Univ. Federal  de Minas Gerais,  
C.P. 702, }\\
{\small \it CEP: 30161-970, Belo Horizonte, M.G., Brazil}\\}

\date{}

\begin{document}

\maketitle

\vspace{-1cm}

\begin{abstract}

The experimental observation of effects due to  Berry's phase 
in quantum systems is certainly one of the 
most impressive demonstrations  of the correctness
of the superposition principle in quantum
mechanics. Since  Berry's original  paper in 1984, the spin
1/2 coupled with rotating external magnetic field  has been 
one of the most studied  models where those phases appear.
 We also consider a special case of this soluble model. 
A detailed analysis of the coupled differential equations
 and comparison with exact results teach us why the usual 
procedure (of neglecting nondiagonal terms) is
mathematically  sound.

\end{abstract}

\vfill

\baselineskip=12pt

\newpage

\baselineskip=18pt

The study of quantum systems in the adiabatic regime appeared for
the first time in the  paper by Born and Fock in 1928\cite{born}.
Since their original paper it is known that the time evolution
of an instantaneous eigenstate of the time-dependent Hamiltonian
acquires a phase besides the dynamical phase.  
But it was only after Berry's seminal paper
in 1984 \cite{berry} that this extra phase was raised to 
the condition of a geometric phase. Different
 quantum systems in this regime have attracted
a lot of attention due to the fact that the existence of those 
geometric phases introduces a shift in the 
frequency  initially predicted  and afterwards  experimentally 
measured. There is a list of experimental articles
on observation of Berry's phase in Ref.\cite{experimental}.
More recently, geometric phases were observed in
 neutrons\cite{neutron} and in an atomic state interacting
with a laser field\cite{atomo}.

Even though the research on quantum systems in the adiabatic
 regime is still a very active area in physics nowadays,
these geometric phases have become an issue included in
undergraduate textbooks in quantum
mechanics\cite{bohm}. Certainly this comes from the
 fact that the measured frequency shift due
to the presence of geometric phases is one of the proofs
 of the linearity  of the quantum phenomena.

The traditional characterization of a quantum system in  
the adiabatic regime is that its interaction with
the environment is described through a set of classical 
time-dependent parameters 
${\bf R}(t) = (R_1(t), \cdots, R_n(t))$ periodic 
in time but whose period is much larger than the
 characteristic time scale involved in the quantum phenomena.
The Hamiltonian that drives
the quantum system depends on the set ${\bf R}(t)$, that is 
${\bf H}[{\bf R}(t)]$. Let $\omega_0$ be
the angular frequency of ${\bf R}(t)$ and $\omega_i$ the
 angular frequencies equal to the difference between 
two distinct quantum eigenvalues of  ${\bf H}[{\bf R}(t)]$  
divided by $\hbar$. The adiabatic regime is 
attained when $\frac{\omega_0}{\omega_i} \rightarrow 0$, for all 
$\omega_i$\cite{berry,bohm,schiff}. Under these conditions, we have 
the so-called the {\it Adiabatic Theorem}\cite{schiff}:

\vspace{0.2cm}
\baselineskip=12pt

{\it For a slowly varing Hamiltonian, the instantaneous
eigenstates of the Hamiltonian evolve continously into
the corresponding eigenstate at a later time.}

\vspace{0.5cm}
\baselineskip=18pt

Let us consider a periodic Hamiltonian that returns to its original
operator, that is, ${\bf H}[{\bf R}(T)] = {\bf H}[{\bf R}(0)]$
when the external classical set ${\bf R}(t)$ completes a period.
If the initial quantum state is an eigenstate of ${\bf H}[{\bf R}(0)]$,
the adiabatic theorem affirms that the evolved state
is an eigenstate of the instantaneous Hamiltonian  with
the same initial quantum number but it is does not
forbid it to acquire a phase. This phase can be decomposed
into a dynamical phase plus another one that M.V. Berry
proved to be geometric\cite{berry}. This  geometrical
phase acquired by instantaneous eigenstates of the Hamiltonian
 chosen as initial state of the quantum system is more
easily recognized if the time evolution
of the initial state  is described in the basis of the
 instantaneous eigenstates of the Hamiltonian
${\bf H}[{\bf R}(t)]$. To make our notation simpler,
we denote: ${\bf H}[{\bf R}(t)] \equiv {\bf H}(t)$.
Let $|\psi(0)\rangle$ be the
initial state of the quantum system; its time evolution
 is given by $|\psi(t)\rangle$. We decompose
$|\psi(t)\rangle$ on the basis of the instantaneous
eigenstates of the Hamiltonian:

\begin{equation}
|\psi(t)\rangle = \sum_{n=1}^{N} c_n(t)
 e^ {-\frac{i}{\hbar} \int_0^t E_n(t^\prime) dt^\prime} \;
 |\phi_n;t\rangle,
        \label{1}
\end{equation}

\noindent where $|\phi_n;t\rangle$ are the instantaneous eigenstates
of ${\bf H}(t)$ with eigenvalue $E_n(t)$,

\begin{equation}
   {\bf H}(t) |\phi_n;t\rangle = E_n(t)  |\phi_n;t\rangle.
     \label{2}
\end{equation}

\noindent The index $n$ represents a set of quantum
numbers that  uniquely determine
 this quantum state.  We are assuming that the time-dependent
 Hamiltonian   has $N$ nondegenerate instantaneous eigenstates.
The phase  $ \; \frac{-i}{\hbar} \int_o^t E_n(t^\prime) dt^\prime \;$
is called the dynamical phase and reduces to
$\frac{-iE_n t}{\hbar}$ when the eigenvalues of the Hamiltonian
are time independent. Requiring  expansion (\ref{1}) to be  a
solution to the Schr\"odinger equation

\begin{equation}
   {\bf H}(t) |\psi(t)\rangle =
     i\hbar \frac{\partial |\psi(t)\rangle}{\partial t},
    \label{3}
\end{equation}

\noindent we get a set of coupled differential equations for the
coefficients $c_n (t)$\cite{berry,bohm,schiff}:

\begin{equation}
\frac{dc_n(t)}{dt}  =
-\langle \phi_n; t|
\frac{\partial }{\partial t}|\phi_n; t\rangle c_n(t)
 + \sum_{m \atop m\not= n}
\frac{ \langle \phi_n; t|
 \frac{\partial {\bf H} (t)}{\partial t}|\phi_m; t\rangle }
{ E_n - E_m}  \;\;
e^{\frac{i}{\hbar} \int_0^t
(E_n(t^\prime) - E_m(t^\prime)) dt^\prime}
    \;c_m(t).        \label{4}
\end{equation}

From Eq.(\ref{1}) up to Eq.(\ref{4}) the expressions are exact. In
Refs. \cite{bohm,schiff}, the adiabatic approximation is
implemented by recognizing that

\begin{equation}
\langle \phi_n; t| \frac{\partial}{\partial t}|\phi_m; t\rangle =
-\frac{\langle \phi_n; t|
 \frac{\partial {\bf H}(t)}{\partial t}|\phi_m; t\rangle}
     {E_n - E_m},
            \hspace{1cm} {\rm for} \hspace{0.5cm} n\not= m.
   \label{5}
\end{equation}

\noindent In this approximation those terms are neglected compared
to the elements
$\langle \phi_n; t| \frac{\partial}{\partial t}|\phi_n; t\rangle$
in Eq.(\ref{4}). After neglecting those terms,
the differential equations (\ref{4}) decouple and each
one is easily solved\cite{bohm,schiff}.

However, since the non-neglected terms in Eq.(\ref{4})
are also a measure of how much the instantaneous eigenstates
 of the Hamiltonian vary with time,  they must also be small once
this variation in time is due to the interaction with the
environment and by hypothesis it is very slow. At first
sight this assertion would seem to contradict the traditional
derivation of the results in the adiabatic
limit\cite{berry,bohm,schiff}.
The aim of this note is to show that this does
not  happen. We explicitly show  this in the context
of a soluble model. In the differential
equations for the coefficients of the expansion of the vector
state [see Eq.(\ref{1})] for this model we introduce tracers  to
 follow the contribution of  crossed terms to the exact solutions
 and finally  we explicitly show  why the adiabatic approximation
 is correctly obtained by only neglecting the terms (\ref{5}).

We consider a particular case of the  model  presented
by M.V. Berry in Ref. \cite{berry}, that is,
a spin $\frac{1}{2}$ coupled to an external
magnetic field $\vec{B}(t)$ with constant norm $B$ ($B \equiv |\vec{B}(t)|$)
that precesses with constant angular frequency $\omega_0$  around
the $z$-axis:

\begin{equation}
\vec{B} (t) = \left( B sin(\theta) cos(\omega_0 t),
B sin(\theta) sin(\omega_0 t), B cos(\theta) \right).
            \label{6}
\end{equation}

\noindent  All directions in space are equivalent, and we assume
that $0\le\theta \le \frac{\pi}{2}$.

The Hamiltonian of the model is

\begin{equation}
H(t) =  \mu \vec{B}(t) \cdot \vec{{\bf s}}.    \label{7}
\end{equation}

\noindent The constant $\mu$ is written in terms
of Land\'e factor $g$  and Bohr magneton $\mu_B$\cite{kittel},
that is, $\mu = g \mu_B$ and $\vec{\bf s}$ is the spin
 $\frac{1}{2}$ operator. Hamiltonian (\ref{7})
is the interaction energy between the intrinsic magnetic
 moment of the electron and the external magnetic
 field\cite{jackson}.  Its instantaneous eigenstates are:

\begin{equation}
|\phi_1; t\rangle = -A |\uparrow\rangle +
           D e^{i\omega_0 t} |\downarrow \rangle
    \hspace{1cm} \Rightarrow \hspace{1cm}
E_1 = -\frac{\mu B\hbar}{2}
      \label{8}
\end{equation}

\vspace{-0.3cm}

\begin{equation}
|\phi_2; t\rangle = D |\uparrow\rangle +
          A e^{i\omega_0 t} |\downarrow \rangle
    \hspace{1cm} \Rightarrow \hspace{1cm}
E_2 = \frac{\mu B\hbar}{2},
      \label{9}
\end{equation}

\noindent where $|\uparrow\rangle$ ($|\downarrow\rangle$) is
the eigenstate of the  component  ${\bf s}_z$ of the spin operator
with eigenvalue $\frac{\hbar}{2}$
($-\frac{\hbar}{2}$). The state
 $|\phi_1; t\rangle$ ($|\phi_2; t\rangle$) is the
eigenvector of spin down (up) along the direction
$\hat{n} = \frac{\vec{B}(t)}{B}$. We defined

\begin{equation}
A\equiv sin (\frac{\theta}{2})
\hspace{1cm} {\rm and} \hspace{1cm}
D\equiv cos (\frac{\theta}{2}).
   \label{10}
\end{equation}

\noindent The eigenvalues $E_1$ and $E_2$ are
time independent. For this model, Eq.(\ref{1}) becomes

\begin{equation}
|\psi(t)\rangle = \sum_{n=1}^{2} c_n(t)
 e^ {-\frac{i}{\hbar} \int_0^t E_n(t^\prime) dt^\prime} \;
     |\phi_n;t\rangle.
        \label{11}
\end{equation}

\noindent The coefficients $c_1(t)$ and $c_2(t)$ satisfy the
differential  equations

\begin{equation}
\frac{dc_1(t)}{dt} = -i \omega_0 D^2 c_1(t)
    - i \omega_0 A D e^{-2i\omega_1t} c_2(t)
   \label{12}
\end{equation}

\vspace{-0.3cm}

\begin{equation}
\frac{dc_2(t)}{dt} = - i \omega_0 A D e^{2i\omega_1t} c_1(t)
-i \omega_0 A^2 c_2(t).
      \label{13}
\end{equation}

\noindent In this model we have only one characteristic
frequency  of the quantum phenomena. It is
proportional to the difference of the eigenvalues
of  Hamiltonian (\ref{7}),

\begin{equation}
2\omega_1 \equiv \frac{E_2 - E_1}{\hbar} = \mu B.
\label{14}
\end{equation}

\noindent The factor $2$ in definition (\ref{14}) is introduced
for convenience. In Eq.(\ref{12}) the adiabatic approximation
($\omega_0\rightarrow 0$) implies that the $c_2(t)$ on its rhs
has to be neglected. However,
the coefficient that multiplies $c_1(t)$ on the rhs of
the same equation  is also proportional to $\omega_0$
although it is not neglected. A similar discussion
is valid for Eq.(\ref{13}) by interchanging $c_1(t)$ and $c_2(t)$.
We should remember that in the adiabatic approximation
 we have no restriction on the $\theta$-angle,
except that it has to be different from $\theta = 0^0$ and
$\theta= \pi$. At first glance it seems that we are not
consistenly keeping the terms in our coupled differential
equations (\ref{12}) and (\ref{13}).
We will see that is not the case and the difference relies on the
exponential that multiplies $c_2(t)$ [$c_1(t)$]  on the rhs
of Eq.(\ref{12}) [Eq.(\ref{13})].

To follow the contribution due to each term on the rhs of
Eqs. (\ref{12}) and (\ref{13}), we introduce tracers
for each term on the rhs in the equations, that is,

\begin{equation}
\frac{dc_1(t)}{dt} = -i a_{11} \omega_0 D^2 c_1(t)
    - i a\omega_0 A D e^{-2i\omega_1t} c_2(t)
   \label{15}
\end{equation}

\vspace{-0.3cm}

\begin{equation}
\frac{dc_2(t)}{dt} = - i a \omega_0 A D e^{2i\omega_1t} c_1(t)
-i a_{22} \omega_0 A^2 c_2(t),
      \label{16}
\end{equation}

\noindent where $a_{11}$, $a_{22}$ and $a$ are the tracers.
These coupled equations are exactly solved.
At the end  we take  the tracers equal to 1 or 0 depending
on the approximation under consideration. The differential
equations in the adiabatic approximation are recovered  by
 chosing $a=0$.  To make the rhs of the equations
time independent, we define new variables, which is equivalent
to going to the rotating frame. The new variables are

\begin{equation}
X_1(t) \equiv c_1(t) e^{i \omega_1 t}
 \hspace{1cm} {\rm and} \hspace{1cm}
X_2(t) \equiv c_2(t) e^{- i \omega_1 t},
 \label{17}
\end{equation}

\noindent and they satisfy the equations

\begin{eqnarray}
\frac{dX_1(t)}{dt}  & = & i (\omega_1 - a_{11} \omega_0 D^2) X_1(t)
      - ia\omega_0 AD X_2(t)  \label{18}  \\
%
%
&  & \nonumber \\
%
%
\frac{dX_2(t)}{dt}  & = &  - ia\omega_0 AD X_1(t)
     - i (\omega_1 + a_{22} \omega_0 A^2) X_2(t).
       \label{19}
\end{eqnarray}

\vspace{0.5cm}

Taking the general initial condition

\begin{equation}
 |\psi(0)\rangle = c_1(0) |\phi_1; 0\rangle +
       c_2(0) |\phi_2; 0\rangle,
         \label{20}
\end{equation}

\noindent where $|\phi_i; 0\rangle$, $i=1,2$, are the
eigenstates of Hamiltonian (\ref{7}) at $t=0$. The exact
solutions of Eqs. (\ref{18}) and (\ref{19}) with
 the initial state (\ref{20}) are

\vspace{-0.7cm}

\begin{eqnarray}
c_1(t) & =&  e^{-i[ \omega_1 + \frac{\omega_0}{2}(a_{11} D^2 +
        a_{22} A^2)]t}
  \left\{ c_1(0) cos (\Gamma t) +    \right.  \nonumber \\
%
%
&  &  \cr
%
%
 &+& \left.
\frac{i\; sin(\Gamma t)}{\Gamma} \left[ (\omega_1 +
 \frac{\omega_0}{2}(a_{22} A^2 - a_{11} D^2))c_1(0) -
            a \omega_0 AD c_2(0)
  \right]
     \right\},
        \label{21}
\end{eqnarray}

\noindent and

\begin{eqnarray}
c_2(t) & =&  e^{i[ \omega_1 - \frac{\omega_0}{2}(a_{11} D^2 +
            a_{22} A^2)]t}
  \left\{ c_2(0) cos (\Gamma t) +    \right.  \nonumber \\
%
%
&  &  \cr
%
%
 &-& \left.
\frac{i\; sin(\Gamma t)}{\Gamma} \left[ (\omega_1 +
 \frac{\omega_0}{2}(a_{22} A^2 - a_{11} D^2))c_2(0) +
          a \omega_0 AD c_1(0)
 \right]
    \right\},
        \label{22}
\end{eqnarray}

\noindent and

\begin{equation}
\Gamma \equiv \omega_1 \left[
\left(1+ \frac{\omega_0}{2\omega_1}(a_{22}A^2 - a_{11} D^2)
      \right)^2  +
a^2 \left(\frac{\omega_0}{\omega_1} \right)^2 A^2 D^2
      \right]^{\frac{1}{2}}.
         \label{23}
\end{equation}

\noindent From Eqs.(\ref{21}) and (\ref{22}) we see that the
nondiagonal terms  in Eqs.(\ref{15}) and (\ref{16})  ($a\not= 0$)
 do not contribute  to the overall
phase in the expressions of $c_1(t)$ and $c_2(t)$. On the other
hand, in the expressions of these coefficients there
appears a new frequency $\Gamma$ (Rabi's frequency\cite{rabi,cohen}).
 For $\omega_1 \not= 0$  and $\omega_1 \gg \omega_0$,
the  contribution of the diagonal
terms  ($a_{11}\not= 0$ and $a_{22}\not= 0$) to the
expression of frequency $\Gamma$ is
 ${\cal O}(\frac{\omega_0}{\omega_1})$ while the contribution
of the nondiagonal terms ($a\not= 0$) is
${\cal O}((\frac{\omega_0}{\omega_1})^2)$.
We get the geometric phase\cite{berry} when the external
parameter  (in our case $\vec{B}(t)$) completes a period
$t= \frac{2\pi}{\omega_0}$\cite{berry} and for this
time scale we can neglect contributions of
${\cal O}((\frac{\omega_0}{\omega_1})^2)$ to the
frequency $\Gamma$. Those are exactly
the  contributions coming from nondiagonal terms in Eqs. (\ref{15})
and (\ref{16}). Due to this fact, we get the adiabatic limit
by substituting  $a=0$ in the expression of $\Gamma$.

The nondiagonal terms in Eqs.(\ref{15}) and (\ref{16}) also
contribute to the {\it sine}-term in Eqs. (\ref{21}) and
(\ref{22}). We now discuss the {\it sine}-term on the rhs
of Eq.(\ref{21}). A similar discussion is valid for
Eq.(\ref{22}). The {\it sine}-term in Eq.(\ref{21}) is
rewritten as

\begin{equation}
 i sin (\Gamma t) \left[ \left(1-
a^2(\frac{\omega_0}{\Gamma})^2 A^2 D^2 \right)^\frac{1}{2} c_1(0)
- a \frac{\omega_0}{\Gamma} AD c_2(0)
\right].
         \label{24}
\end{equation}

\noindent In the adiabatic regime we have
$\frac{\omega_0}{\omega_1} \ll 1$ and then from
Eq.(\ref{23}) it follows that $\frac{\omega_0}{\Gamma} \ll 1$.
Keeping terms up to zeroth order in
$\frac{\omega_0}{\Gamma} \ll 1$ in the {\it sine}-term (\ref{24}),
we obtain

\begin{equation}
i sin (\Gamma t) \left[ \left(1-
a^2(\frac{\omega_0}{\Gamma})^2 A^2 D^2 \right)^\frac{1}{2} c_1(0)
- a \frac{\omega_0}{\Gamma} AD c_2(0)
\right] \approx i c_1(0) sin(\Gamma t).
    \label{25}
\end{equation}

\noindent Substituting the approximation (\ref{25}) in the
expression of $c_1(t)$,  we get in the adiabatic limit

\begin{equation}
c_1^{a.a.} (t) = exp\left[- i\left( \omega_1 +
\frac{\omega_0}{2} (a_{11} D^2 + a_{22} A^2) -
   \Gamma^{a.a.}\right) t \right]  c_1(0),
     \label{26}
\end{equation}

\noindent with

\begin{equation}
\Gamma^{a.a.} = \omega_1 \left[ 1 + \frac{\omega_0}{2 \omega_1}
( a_{22} A^2 - a_{11} D^2)   \right],
     \label {27}
\end{equation}

\noindent that is, the Rabi's frequency in the adiabatic
approximation.

By the same procedure we obtain, in the adiabatic regime,

\begin{equation}
c_2^{a.a.} (t) = exp\left[ i\left( \omega_1 -
\frac{\omega_0}{2} (a_{11} D^2 + a_{22} A^2) -
   \Gamma^{a.a.}\right) t \right]   c_2(0).
     \label{28}
\end{equation}

\noindent Taking $a_{11} = a_{22} = 1$, in Eqs.(\ref{26}) and
(\ref{28}) we recover the known results for the adiabatic
approximation\cite{berry,bohm}, which is equivalent to making
$a=0$ in Eqs.(\ref{15}) and (\ref{16}) from the beginning.

\vspace{0.5cm}

We showed analitically, for a two-level model, why we get the
correct adiabatic limit expressions by just neglecting the
nondiagonal terms in the coupled equations (\ref{12})
and (\ref{13}). These first-order differential equations
drive the time evolution of the expansion coefficients
(\ref{1}) for the quantum state on the basis of the
instantaneous eigenstate of the time-dependent Hamiltonian.

It certainly could be questioned why terms of different orders
in $(\frac{\omega_0}{\omega_1})$ are dropped in the expansions of
the coefficients $c_i(t)$, $i=1,2$, and of the frequency $\Gamma$. The
big difference between these two expansions  is due to the fact
that we are interested in time scale of order
$T= \frac{2\pi}{\omega_0}$, where the linear correction in
$(\frac{\omega_0}{\omega_1})$  to $\Gamma$ gives a
finite contribution. This does not happen to the linear correction
$(\frac{\omega_0}{\omega_1})$  to the amplitude of
the terms $\frac{sin(\Gamma t)}{\Gamma}$.

Even though we studied a soluble spin $\frac{1}{2}$ model,
many features of this model appear in any quantum system that
 is coupled   to a tridimensional external parameter
 that varies in time  very slowly. For these models we can have a
matrix representation of Eq.(\ref{3}) on a suitable basis
such that the coefficients can
be written as a column where the equations that drive their 
time evolution are a generalization of Eqs.(\ref{12})
and (\ref{13}). Neglecting  the nondiagonal
terms in the adiabatic  limit is justified  if they are
 multiplied by a phase whose angular velocity is much larger
than the angular frequency of the external parameter.

\vspace{0.5cm}

\section*{\bf Acknowledgements}

A.C.A.P. and J.G.P.F. thank CNPq for financial support. M.C.N. thanks
 CNPq for partial financial support.  M.T.T.  thanks 
CNPq and  FAPERJ for partial financial suport.

\end{document}